\documentclass[usenatbib]{mnras}
\usepackage[utf8]{inputenc}
\usepackage{natbib,epsfig,cite,amsmath,amssymb,graphicx}
\usepackage{array, xcolor, caption, multirow}

\def\e10{\eta_{10}}

\def\iso#1#2{\mbox{${}^{#2}{\rm #1}$}}

\def\b1#1{\iso{B}{1#1}}
\def\msun{\mbox{$M_\odot$}}

\def\beq{\begin{equation}}
\def\eeq{\end{equation}}
\def\beqar{\begin{eqnarray}}
\def\eeqar{\end{eqnarray}}
\def\simlt{\lower.5ex\hbox{$\; \buildrel < \over \sim \;$}}
\def\simgt{\lower.5ex\hbox{$\; \buildrel > \over \sim \;$}}
\def\simpropto{\lower.2ex\hbox{$\; \buildrel \propto \over \sim \;$}}

\title[Cosmic Evolution of Magnesium]{The Cosmic Evolution of Magnesium Isotopes}
 
\author[E. Vangioni and K.A. Olive]
{Elisabeth Vangioni$^{1}$\thanks{e-mail:vangioni@iap.fr},
Keith A. Olive$^{2}$\\
$^{1}$Sorbonne Universit\'e, UPMC Univ Paris 6 et CNRS, UMR 7095, Institut d'Astrophysique de Paris, 98 bis bd Arago, 75014 Paris, France\\
$^{2}$William I. Fine Theoretical Physics Institute, School of Physics and Astronomy, 
University of Minnesota, Minneapolis, MN 55455, USA\\
}
 
\begin{document}

\pagerange{\pageref{firstpage}--\pageref{lastpage}} \pubyear{2018}
\maketitle
\label{firstpage}

\begin{abstract}
The abundance of magnesium in the interstellar medium is a powerful probe of star formation processes over cosmological timescales.  Magnesium has three stable isotopes, $^{24}$Mg, $^{25}$Mg, $^{26}$Mg, which can be produced both in massive and intermediate-mass (IM) stars with masses between 2 and 8 M$_\odot$. In this work, we use
constraints on the cosmic star formation rate density (SFRD) and explore the role and 
mass range of intermediate mass stars
using the observed isotopic ratios. 
We compare several models of stellar nucleosynthesis with  
metallicity-dependent yields and also consider the effect of rotation on the yields massive stars and its consequences
on the evolution of the Mg isotopes. We use a cosmic evolution model updated with new observational SFRD data and new reionization constraints coming from 2018 Planck collaboration determinations. 
We find that the main contribution of $^{24}$Mg comes from massive stars whereas $^{25}$Mg and $^{26}$Mg come from intermediate mass stars. To fit the observational data on magnesium isotopic ratios, an additional intermediate mass SFRD component is preferred. Moreover, the agreement between model and data is further improved when the range of
IM masses is narrowed towards higher masses (5-8 M$_\odot$). While some rotation also improves the fit to data,
we can exclude the case where all stars have high rotational velocities due to an over-production of $^{26}$Mg.

\end{abstract}


\section{Introduction}
One of the most powerful probes of the baryonic history of the Universe  is the determination of the elemental and isotopic abundances in both the interstellar medium (ISM)
and the Universe at high redshift. 
The initial conditions are set by big bang nucleosynthesis \citep{2016RvMP...88a5004C,2018arXiv180108023P} giving 25\% of the baryonic mass in $^{4}$He,
and a deuterium abundance of $2.5 \times 10^{-5}$ by number (with a similar abundance for $^{3}$He and a much smaller abundance of $5 \times 10^{-10}$ for
$^{7}$Li). All of the remaining element abundances are the result of the star formation history which can be complicated by gas accretion and galactic outflows. 
Galactic chemical evolution models \citep{1972A&A....20..383T,1975MNRAS.172...13P,1978ApJ...221..554T,2004oee..symp...85M,2006ApJ...653.1145K} 
have been successful in describing the chemical enrichment of our own Galaxy, but 
to understand the chemical history of the Universe at high redshift, we must turn to cosmic chemical evolution models
\citep[e.g.][]{daigne06,2012MNRAS.421...98D,2013ApJ...772..119L,2015ApJ...808..129L,2016MNRAS.455.1218B}.

These models, however, are developed using many parameters whose values can only be inferred from observations.
For example, all such models require an initial mass function (IMF) and star formation rate (SFR).
The latter can be taken from observations of the luminosity function as a function of redshift \citep{1996ApJ...460L...1L,Madau14, Behroozi13,Bouwens14, O14,  O18}. 
The IMF on the other hand requires several assumptions.
Is there a unique IMF? Does the IMF vary with time (or redshift)? What are the minimum and maximum masses of stars which participate in gas consumption,
and what is the minimum and maximum masses of stars which participate in element production?
To what extent does infall or outflows of gas play a role? What is the history of baryons in star forming structures?
Of course the parameters associated with these questions may be degenerate as well with different assumptions leading to similar
observables. Finally (though not exhaustively), how well are the yields of massive stars known as function
of mass, metallicity and other intrinsic properties such as rotation?
Remarkably, despite definitive answers to all of these questions, a general framework for the star formation
history can be developed and 
observations of individual element abundances, as well as their relative abundances are particularly informative. 
These observations can be used to constrain the nucleosynthetic history as well as the specific mass ranges of stars responsible for their production.

Here, we study in detail, the production and evolution of the magnesium isotopes, $^{24}$Mg, $^{25}$Mg, and $^{26}$Mg.
 The observational determination of these isotopic abundances has been given in many studies 
 \citep{1968ApJ...154..185B, 1970ApL.....5..203B, 1976ApJ...208..436T, 1980ApJ...235..925T,1985A&A...151..189B,Barbuy87, 1988MNRAS.230..573M, Shetrone96, Gay00, Yong03a, Yong03b, Yong06, Melendez07, Melendez09, Agafonova11, Yong13, Dacosta13, Thygesen16}.
  Magnesium is an interesting element because its different isotopes are produced in different sites.  The isotopes $^{24,25,26}$Mg are produced inside massive stars but $^{24}$Mg is the most abundant, and we confirm below that the total budget of this element (isotope $^{24}$Mg) is dominated by massive stars in star-forming galaxies during core carbon and neon burning before the supernova explosion  \citep{Heger10}.
 The isotopes $^{25,26}$Mg are predominantly produced in stars with intermediate mass (IM) \citep{Karakas14}
in the outer carbon layer through $\alpha$ capture on neon. 
Consequently, these Mg isotopes which originate from asymptotic giant branch (AGB) stars begin to contribute later as galactic chemical enrichment evolves. 
That is, the isotopic ratios are expected to increase as a function of the enrichment or metallicity.
However, as we will see,  it is presently difficult to explain this increase in the context of
 a standard evolutionary model. 
 As a consequence, we can use the magnesium
 isotope ratios to constrain the role of an intermediate mass stellar component in cosmic chemical evolution.

The chemical evolution of the magnesium isotopes in the ISM was extensively studied in the context of galaxy evolution models \citep{Gay00, 2001A&A...370.1103A,
Fenner03, Ashen04,  2004ApJ...615...82A, Koba10,  Thygesen16, Thygesen17, Carlos18}.
The yield of the heavier isotopes scales
 with the metallicity in the carbon layer and as a result, very little of these isotopes are produced
 at low metallicity. 
 In contrast, significant amounts of $^{25,26}$Mg are produced during hot bottom-burning in the AGB phase in intermediate mass stars \citep{booth}.  
 It should be noted that there is an inherent uncertainty in chemical evolution models stemming from the uncertainty in theoretical
stellar abundances \citep{hg97,sll,2003PASA...20..279K, denher,Karakas10, dohertyII, dohertyIII, Karakas14,vetal}. 
Nevertheless, all of these studies point to the fact that AGB stars are required to fit isotope observations at late times.
 
 The importance of an IM component to the IMF is accentuated by 
 several observations \citep{Shetrone96, Yong03a, Yong03b, Agafonova11, Dacosta13, webb14} which indicate enhancements 
 over terrestrial abundances of the
 neutron-rich isotopes in low metallicity stars which could necessitate the presence of an
 early population of intermediate mass stars \citep{2001A&A...370.1103A, Fenner03, Ashen04, 
2004ApJ...615...82A}. A substantial IM component producing large abundances of the heavy isotopes was considered by \citet{Ashen04, 2004ApJ...615...82A}, to study the impact on claims for a potential variation of fundamental constants. Such models can be constrained by the nitrogen abundance at high redshift \citep{Fenner05}. 
A detailed study of the impact of cosmic chemical evolution on nitrogen was recently performed in \citet{Vangioni18}.
In fact, \citet{Agafonova11}  has made a measurement of the Mg isotopic abundances at high redshift to deduce 
an estimate of a variation of the fine-structure constant, 
$\alpha$.  \citet{webb14} also find that $^{24}$Mg is suppressed while $^{25,26}$Mg is enhanced
in high redshift absorbers.

 In addition to the magnesium isotopes, 
there are several motivations for including an intermediate mass mode in cosmic chemical evolution models.
First of all, these stars may be part of a secondary population of early stars (PopIII.2) which originated
from material polluted by zero-metallicity 
PopIII (or PopIII.1) stars \citep{bromm2}. 
There are also theoretical arguments that the zero-metallicity IMF predicted from opacity-limited 
fragmentation theory should peak around 4 -- 10 M$_\odot$ with steep declines
at both larger and smaller masses \citep{yoshii}.  
Primordial CMB regulated-star formation  may also lead to the
production of a population of early intermediate mass stars  at low metallicity \citep{tumlinson,smith09, schneider10}. 

There is also some evidence for an early contribution by IM stars from observations. 
While these stars produce only a small fraction of the total abundance of heavy elements (oxygen and above),
they do produce significant amounts of helium, carbon and/or nitrogen as indicated above.
Contributions to the helium and CNO abundances from a population of IM stars was considered in \citet{vsof}.
There is evidence that the number of carbon-enhanced metal-poor stars
increases at low iron abundances \citep{rossi} necessitating a PopIII
source of carbon. While the source of this carbon is uncertain, there is the possibility that
its origin lies in the AGB phase of 
IM stars \citep{fujimoto,aoki2,lucatello1,tumlinson2} indicating possibly  an IMF peaked at
4 - 10 M$_\odot$  \citep{abia}.
Finally, the presence of s-process elements, 
at very low metallicities also points to an AGB enrichment very early on \citep{aoki,sivarani,lucatello2}.

 In order to track the evolution of the magnesium isotopes, we work in the context of a cosmic chemical evolution model 
\citep{daigne04, daigne06, 2009MNRAS.398.1782R, vangioni15}. Our base model, employs a Salpeter IMF and a SFR as a function of redshift fit to the 
observed SFR density (SFRD) derived from the luminosity function. To assess the impact and importance of a population of IM stars we consider a bimodal IMF \citep{larson86,ws87,vfa}
where the second mode of star formation consists of IM stars between 2 and 8 M$_\odot$. 
Such models are known to affect the evolution of chemical abundances, particularly, light elements \citep{vop,1997ApJ...476..521S,cova,vsof,vangioni15,Vangioni18}.
In addition, galactic bimodal models were utilized in attempted resolution of the g-dwarf problem \citep{ogdwarf,fvfa}.

An argument for the presence of an additional component of IM stars has also been made on the basis of 
a conflict between extra-galactic background light (EBL) density and the K-band light density \citep{fardal}. 
Using a standard (or single sloped) IMF typically produces either a deficit of EBL density or an excess in
the K-band light density. This discrepancy may be resolved with an excess (over a simple single sloped IMF)
of IM stars with masses between 1 and 4 M$_\odot$.
Recent observations of the UV-to-mm extragalactic background may continue to point to a bimodal mass function \citep{2018arXiv180805208C} and a preferred excess of stars with masses between 1 and 8 M$_\odot$.

To judge the robustness of our results, 
we consider different sets of chemical yields, including the effect of the rotation of massive stars. 
 The goal of the current study is to examine the uncertainties in different stellar evolution models and their effect on the predicted cosmic evolution of the magnesium isotope abundances. 

The paper is organized as follows: in Section~\ref{sec:nucleosynthesis}, we review the production of magnesium in different stellar evolution models. In particular, we discuss the dependence on stellar mass, metallicity, and rotation velocity. In Section~\ref{sec:meanevolution}  we describe the cosmic evolutionary model. In Section~\ref{sec:CNO} we show how these yields are implemented and we calculate the mean evolution of the CNO and Mg isotope abundances in the interstellar medium. We conclude in Section~\ref{sec:conclusion}.

\section{Stellar nucleosynthesis of Magnesium}
\label{sec:nucleosynthesis}

In any chemical evolution model, galactic or cosmic, the abundances as a function of time or redshift depend
critically on the calculated yields from massive and intermediate mass stars. 
Much of the total yield of magnesium arises from massive stars exploding as type II supernovae. 
However, the heavy isotopes receive substantial contributions from the AGB phase of intermediate mass stars.
We consider these two populations separately in the analysis below. We also consider the effect of stellar rotation
on the yields.

\subsection{Massive stars}
\label{sec:massivestars}

 We employ the results of three different stellar evolution models.
 Our primary results are based on the supernova yields calculated in \citet{Nomoto06}.
 \citet{Nomoto06} study the mass range $13-35M_\odot$ for $4$ different metallicities: $Z= 0, 10^{-3}, 0.004, 0.02$. 
 We interpolate the yields at intermediate metallicities. 
 For the purposes of comparison and as a check on the robustness of our results, we also consider the yields of 
\citet{ww95}  and \citet{Limongi18}.
 \citet{ww95} present the evolution of massive stars at $5$ different metallicities ($Z/Z_{\odot}=0, 10^{-4}, 10^{-2}, 0.1, 1$) and masses from $11M_{\odot}$ to $40M_\odot$.
\citet{Limongi18} provide another set of explosive yields for masses in the range $13 - 120 M_\odot$ at $4$ different metallicities ($[Fe/H]= 10^{-3},10^{-2}, 10^{-1}, 0$). 
 For illustration, Table \ref{table1} compares typical yields produced by these models at two metallicities ($Z/Z_{\odot}=0.1, 1$) and two masses: $15M_{\odot},30M_{\odot}$.

\begin{table*}
\caption{Typical magnesium isotope yields of massive stars as a function of stellar mass and initial metallicity in different stellar evolution models (masses are in solar mass units):  WW95: \citet{ww95},  Nomoto06: \citet{Nomoto06}, LCH18: \citet{Limongi18},  without rotational effect.}
\begin{center}
\begin{tabular}{c|c|c|c|c|c|c}
 Metallicity  & \multicolumn{3}{c|}{$10^{-1}Z_{\odot}$} & \multicolumn{3}{c|}{$10^{-1}Z_{\odot}$} \\
\hline
 Stellar Mass &  \multicolumn{3}{c|}{15} & \multicolumn{3}{c|}{30}\\
\hline 
 Element &  $^{24}Mg$ &   $^{25}Mg$  &   $^{26}Mg$  &   $^{24}Mg$  &   $^{25}Mg$  &   $^{26}Mg$    \\
\hline 
WW95  & $2.2\cdot 10^{-2}$ &  $6.0\cdot 10^{-4}$ & $5.1\cdot 10^{-4}$ &  $3.4\cdot 10^{-1}$ & $3.8\cdot 10^{-3}$ & $4.4\cdot 10^{-3}$ \\
Nomoto06 & $6.4\cdot 10^{-2}$ &$ 8.8\cdot 10^{-4}$ & $1.1\cdot 10^{-3}$ & $2.9\cdot 10^{-1}$ & $3.6\cdot 10^{-3}$ & $4.3\cdot 10^{-3}$ \\
LCH18 & $3.1\cdot 10^{-2}$ & $1.1\cdot 10^{-3}$ & $1.1\cdot 10^{-3}$ & $1.7\cdot 10^{-4}$ & $2.2\cdot 10^{-5}$ & $2.6\cdot 10^{-5}$ \\
\hline
\end{tabular}
\end{center}

\begin{center}
\begin{tabular}{c|c|c|c|c|c|c}
 Metallicity  &  \multicolumn{3}{c|}{$Solar$} & \multicolumn{3}{c}{$Solar$} \\
\hline
 Stellar Mass &  \multicolumn{3}{c|}{15} & \multicolumn{3}{c|}{30}\\
\hline 
 Element &  $^{24}Mg$  &   $^{25}Mg$  &   $^{26}Mg$  &   $^{24}Mg$  &   $^{25}Mg$  &   $^{26}Mg$  \\
\hline 
WW95  &  $2.7\cdot 10^{-2}$ & $6.7\cdot 10^{-3}$ & $6.5\cdot 10^{-3}$ & $2.8\cdot 10^{-1}$ & $2.9\cdot 10^{-2}$ & $3.6\cdot 10^{-2}$\\
Nomoto06 & $3.8\cdot 10^{-2}$ & $1.5\cdot 10^{-3}$ & $1.7\cdot 10^{-3}$ &$1.9\cdot 10^{-1}$ & $3.1\cdot 10^{-2}$ & $7.3\cdot 10^{-2}$\\
LCH18 &  $3.9\cdot 10^{-2}$ & $4.1\cdot 10^{-3}$ & $4.2\cdot 10^{-3}$ & $1.1\cdot 10^{-2}$ & $9.8\cdot 10^{-4}$ & $2.0\cdot 10^{-3}$\\
\hline
\label{table1}
\end{tabular}
\end{center}
\end{table*}

\subsection{Intermediate-mass stars}
\label{sec:imstars}

As noted earlier, the $^{25,26}$Mg isotopes are produced during hot bottom-burning in the AGB phase in intermediate mass stars \citep{booth}.  
 These stars are hot enough
 for efficient proton capture processes on Mg leading to Al (which decays to the heavier Mg isotopes).
 The neutron-rich isotopes are also produced during thermal pulses of the helium burning shell. 
 Here, $\alpha$ captures on $^{22}$Ne (which is produced from $\alpha$ capture on $^{14}$N)
 lead to both $^{25}$Mg and $^{26}$Mg.

We compare the results of two different stellar evolution models that target the evolution of intermediate-mass stars. \citet{hg97} provide the evolution of stars at $5$ metallicities ($Z = 0.001, 0.004, 0.008, 0.02, 0.04$) with masses in the range $0.8- 8M_\odot$. \citet{Karakas10} provides an independent set of yields for masses in the range $1-6M_\odot$ at $4$ different metallicities ($Z= 0.0001, 0.004, 0.008, 0.02$). Additional yields for $Z = 0.001$ and the same mass range are found in \citep{Fishlock14}.  
Table \ref{table2} presents typical yields for two metallicities ($Z/Z_{\odot}=0.1, 1$)  and two masses, $2M_{\odot}$ and $7M_\odot$.

\begin{table*}
\caption{Typical magnesium isotope yields of intermediate-mass stars as a function of stellar mass and initial metallicity in different stellar evolution models (masses are in solar mass units): VdH97: \citet{hg97}, Karakas10: \citet{Karakas10}.}
\begin{center}
\begin{tabular}{c|c|c|c|c|c|c}
Metallicity &  \multicolumn{3}{c|}{$10^{-1}Z_{\odot}$} & \multicolumn{3}{c|}{$10^{-1}Z_{\odot}$} \\
\hline
Stellar Mass & \multicolumn{3}{c|}{2} & \multicolumn{3}{c|}{7}  \\
\hline 
 Element &  $^{24}Mg$ &   $^{25}Mg$  &   $^{26}Mg$  &   $^{24}Mg$  &   $^{25}Mg$  &   $^{26}Mg$   \\
\hline
VdH97  & $2.7\cdot 10^{-3}$ & $3.5\cdot 10^{-5}$ & $4.1\cdot 10^{-5}$ & $1.3\cdot 10^{-4}$ & $1.4\cdot 10^{-3}$ & $8.5\cdot 10^{-4}$ \\
Karakas10 & $1.9\cdot 10^{-5}$ & $4.9\cdot 10^{-5}$ & $5.6\cdot 10^{-5}$ & $2.0\cdot 10^{-5}$ & $5.5\cdot 10^{-4}$ & $5.0\cdot 10^{-4}$ \\
\hline
\end{tabular}
\end{center}

\begin{center}
\begin{tabular}{c|c|c|c|c|c|c}
Metallicity &  \multicolumn{3}{c|}{$Solar$} & \multicolumn{3}{c}{$Solar$}\\
\hline
Stellar Mass & \multicolumn{3}{c|}{2} & \multicolumn{3}{c|}{7} \\
\hline 
 Element &  $^{24}Mg$ &   $^{25}Mg$  &   $^{26}Mg$  &   $^{24}Mg$  &   $^{25}Mg$  &   $^{26}Mg$   \\
\hline
VdH97  &  $7.4\cdot 10^{-4}$ & $9.8\cdot 10^{-5}$ & $1.1\cdot 10^{-4}$ & $2.9\cdot 10^{-4}$ & $7.0\cdot 10^{-4}$ & $8.3\cdot 10^{-4}$ \\
Karakas10 & $7.0\cdot 10^{-4}$ & $9.2\cdot 10^{-5}$ & $1.1\cdot 10^{-4}$ & $2.9\cdot 10^{-3}$ & $7.0\cdot 10^{-4}$ & $8.3\cdot 10^{-4}$ \\
\hline
\label{table2}
\end{tabular}
\end{center}
\end{table*}

\subsection{Stellar rotation}
\label{sec:rotation}

Stellar rotation, particularly important at low metallicities, can strongly affect the nucleosynthetic yields \citep{mm02a, mm02b, meynet04}. 
Table \ref{table3} compares the yields calculated by \citet{Limongi18} of rotating ($v=150$ and $v= 300$ km/s) and non-rotating stars  for three metallicities ($[Fe/H]= 0.01, 0.1, 1$)  and two masses, $15M_{\odot}$ and $30M_\odot$. 
As can be seen from the table, the Mg isotopes  are produced in larger amounts in rotating stars, and the difference with the non-rotating case can reach an order of magnitude.

\begin{table*}
\caption{Typical magnesium isotope yields (in solar mass units) of rotating and non-rotating stars for different stellar masses and metallicities , \citet{Limongi18}.}

\begin{center}
\begin{tabular}{c|c|c|c|c|c|c}
 Metallicity &  \multicolumn{3}{c|}{$10^{-2}$ [Fe/H]} & \multicolumn{3}{c|}{$10^{-2}$ [Fe/H]}\\
\hline
 Stellar Mass & \multicolumn{3}{c|}{15} & \multicolumn{3}{c|}{30}  \\
\hline 
 Element &  $^{24}Mg$ &   $^{25}Mg$  &   $^{26}Mg$  &   $^{24}Mg$  &   $^{25}Mg$  &   $^{26}Mg$  \\
\hline
$v=0$ km/s & $2.5\cdot 10^{-2}$ & $2.0\cdot 10^{-4}$ & $2.5\cdot 10^{-4}$ & $1.3\cdot 10^{-6}$ & $1.7\cdot 10^{-7}$ & $2.0\cdot 10^{-7}$\\
$v=150$ km/s &  $1.5\cdot 10^{-2}$ & $1.3\cdot 10^{-3}$ & $8.6\cdot 10^{-4}$ & $3,2\cdot 10^{-5}$ & $2.5\cdot 10^{-6}$ & $4.0\cdot 10^{-6}$\\
$v=300$ km/s & $1.0\cdot 10^{-2}$ & $1.8\cdot 10^{-2}$ & $4.4\cdot 10^{-2}$ & $1.5\cdot 10^{-4}$ & $1.1\cdot 10^{-5}$ & $1.8\cdot 10^{-5}$\\
\end{tabular}
\end{center}

\begin{center}
\begin{tabular}{c|c|c|c|c|c|c}
 Metallicity &  \multicolumn{3}{c|}{$10^{-1}$ [Fe/H]} & \multicolumn{3}{c|}{$10^{-1}$ [Fe/H]} \\
\hline
 Stellar Mass & \multicolumn{3}{c|}{15} & \multicolumn{3}{c|}{30}  \\
\hline 
 Element &  $^{24}Mg$ &   $^{25}Mg$  &   $^{26}Mg$  &   $^{24}Mg$  &   $^{25}Mg$  &   $^{26}Mg$   \\
\hline
$v=0$ km/s & $3.0\cdot 10^{-2}$ & $1.0\cdot 10^{-3}$ & $1.0\cdot 10^{-3}$ & $1.7\cdot 10^{-4}$ & $2.2\cdot 10^{-5}$ & $2.5\cdot 10^{-5}$\\
$v=150$ km/s &$3.4\cdot 10^{-2}$ & $ 2.9\cdot 10^{-3}$ & $2.9\cdot 10^{-3}$ & $ 4.0\cdot 10^{-4}$ & $3.4\cdot 10^{-5}$ & $6.9\cdot 10^{-5}$ \\
$v=300$ km/s & $7.4\cdot 10^{-3}$ & $1.4\cdot 10^{-2}$ & $3.1\cdot 10^{-2}$ & $1.3\cdot 10^{-3}$ & $8.9\cdot 10^{-5}$ & $2.4\cdot 10^{-4}$ \\
\end{tabular}
\end{center}

\begin{center}
\begin{tabular}{c|c|c|c|c|c|c}
 Metallicity &  \multicolumn{3}{c|}{$Solar$} & \multicolumn{3}{c}{$Solar$}\\
\hline
 Stellar Mass & \multicolumn{3}{c|}{15} & \multicolumn{3}{c|}{30}  \\
\hline 
 Element &  $^{24}Mg$ &   $^{25}Mg$  &   $^{26}Mg$  &   $^{24}Mg$  &   $^{25}Mg$  &   $^{26}Mg$  \\
\hline
$v=0$ km/s & $3.8\cdot 10^{-2}$ & $4.1\cdot 10^{-3}$ & $4.1\cdot 10^{-3}$ & $1.0\cdot 10^{-2}$ & $9.8\cdot 10^{-4}$ & $1.9\cdot 10^{-3}$\\
$v=150$ km/s &  $1.6\cdot 10^{-2}$ & $1.3\cdot 10^{-2}$ & $1.0\cdot 10^{-2}$ & $1.0\cdot 10^{-2}$ & $6.2\cdot 10^{-4}$ & $2.1\cdot 10^{-3}$\\
$v=300$ km/s & $2.3\cdot 10^{-2}$ & $1.1\cdot 10^{-2}$ & $1.2\cdot 10^{-2}$ & $1.0\cdot 10^{-2}$ & $4.6\cdot 10^{-4}$ & $2.2\cdot 10^{-3}$\\
\label{table3}
\end{tabular}
\end{center}
\end{table*}

\section{Cosmic chemical evolution model}
\label{sec:meanevolution}

While galactic chemical evolution models often include the effect of infall and outflow, 
cosmic chemical evolution model in addition track the growth of star forming structures and the baryon fraction found in these structures. 
Here, we use the models developed by \citet{daigne04, daigne06}, \citet{2009MNRAS.398.1782R} and \citet{vangioni15} to investigate the
evolution of the magnesium isotopes. 
The mean baryon fraction, $f_{baryon}=\Omega_b/\Omega_m$, where $\Omega_b$ and $\Omega_m$ 
are the densities of baryons and total dark matter, respectively, in units of the critical density of the Universe, determines the initial baryon fraction within a galaxy. 
 However,  in a model of hierarchical structure formation, the mean baryon accretion rate in each region is proportional to the fraction of 
baryons in structures, $f_{coll}$, and can be expressed as 
\begin{equation}
a_\mathrm{b}(t)  =  \Omega_\mathrm{b}\left(\frac{3H_{0}^{2}}{8\pi G}\right)\ \left(\frac{dt}{dz}\right)^{-1}\ \left|\frac{d f_{coll}}{dz}\right|
\end{equation}
where $f_{coll}(z)$ is given by \citep{ps74}, 
\begin{equation}
f_{coll}(z) = \frac{\int_{M_\mathrm{min}}^{\infty} dM\ M f_\mathrm{PS}(M,z)}{\int_{0}^{\infty} dM\ M f_\mathrm{PS}(M,z)}\ .
\label{eq:fb}
\end{equation}
The model assumes that the minimum mass of dark matter haloes for star-forming structures is $10^{7}~M_\odot$. 
The gas is assumed initially to be primordial and metal-free and we begin the calculation at a redshift $z = 20$. 

The model contains two gas reservoirs corresponding to intergalactic matter (IGM) and the ISM.
Accreted baryons flow from the IGM to the ISM as galaxies form. 
Within a galaxy, baryons form stars at a rate $\psi(t)$ which is fit to observations of the cosmic SFRD. 
We assume a Salpeter IMF, $\Phi(m)$, with slope $x = 1.35$, for $m_{inf}\leq m \leq m_{sup}$ with $m_{inf}=0.1 M_{\odot}$ and $m_{sup}=100M_{\odot}$. 
Due to galactic winds and outflows, 
baryons can also flow from structures to the IGM. 
As a consequence, we can track the baryon content of both the IGM and the ISM using
\begin{equation}
\frac{dM_{IGM}}{dt}=-a_b(t)+o(t)
\end{equation}
and 
\begin{equation}
\frac{dM_{ISM}}{dt}=-\psi(t)+e(t)+a_b(t)-o(t),
\end{equation}
where $e(t)$ is the rate at which baryons are returned to the ISM by mass loss or stellar deaths and $o(t)$ is the baryon outflow rate from structures into the IGM. 

The outflow rate is assumed to have two components \citep{daigne04} which differ in their chemical composition.
One component is effectively a galactic wind powered by supernova explosions and is similar to that described in
\citet{1997ApJ...476..521S}.  The composition of this component is the same as the ISM.
The second component is metal enriched compared to the ISM and corresponds to the fraction
 of stellar supernova ejecta which is flushed directly out of the
 structures \citep{1986ApJ...305..669V}.  

In order to avoid the instantaneous recycling approximation, we 
include the effect of stellar lifetimes.
The lifetimes of intermediate mass stars ($0.9<M/\msun <8$) are taken
 from \citet{mm89} and  from
\citet{schaerer02} for more massive stars. 
Further details on the chemical evolution model can be found in \citet{daigne04,daigne06} and \citet{vangioni15}.

Over the last decade there have been significant improvements in the observations of the SFRD out to high redshift, $z \sim 10$. The SFRD
is quite well measured out  to $z \sim 3$. We use the observations compiled by \citet{Behroozi13}, augmented by observations at higher redshift in \citet{Bouwens14, O14, O18}.
The model assumes an analytic form for the SFRD given by \citep{springel03}
\begin{equation}
\psi(z) = \nu\frac{a\exp(b\,(z-z_m))}{a-b+b\exp(a\,(z-z_m))}
\end{equation}
with parameters fit directly to the observed SFRD \citep{2009MNRAS.398.1782R,vangioni15}. 
\citet{vangioni15} found a good fit with $\nu = 0.178 $  M$_{\odot}$/yr/Mpc$^{3}$, 
$z_m = 2.00, a = 2.37$ and $b=1.80$. 
Taking into account the new observational constraints given by \citet{O18},
here, we take  $b= 1.69$.
Indeed, these authors show a decrease in the SFRD between $z = 8$ to $z = 10$. 
This is our base model which we will refer to as Model 1.

To account for the observed abundances of the heavy magnesium isotopes, we
will also consider a bimodal model where the 2nd mode of star formation consists of IM stars.
For the second mode, we use the same form for the SFRD, but with 
$\nu$ = 0.1 or 0.05   M$_{\odot}$/yr/Mpc$^{3}$, 
$z_m = 1.00, a = 2.37, b=1.69$. 
 We assume a Salpeter IMF, $\Phi(m)$, with slope $x = 1.35$, for $m_{inf}\leq m \leq m_{sup}$ with $m_{inf}=2 M_{\odot}$ (Model 2a) or $m_{inf}=5 M_{\odot}$  (Model 2b) and $m_{sup}=8M_{\odot}$. This component, added to Model 1 will be denoted Model 2. Since observations are now available out to $z = 10$, the SFRD is highly constrained and severely
 limits any additional component to the SFRD. 

\begin{figure}
\begin{center}
\epsfig{file=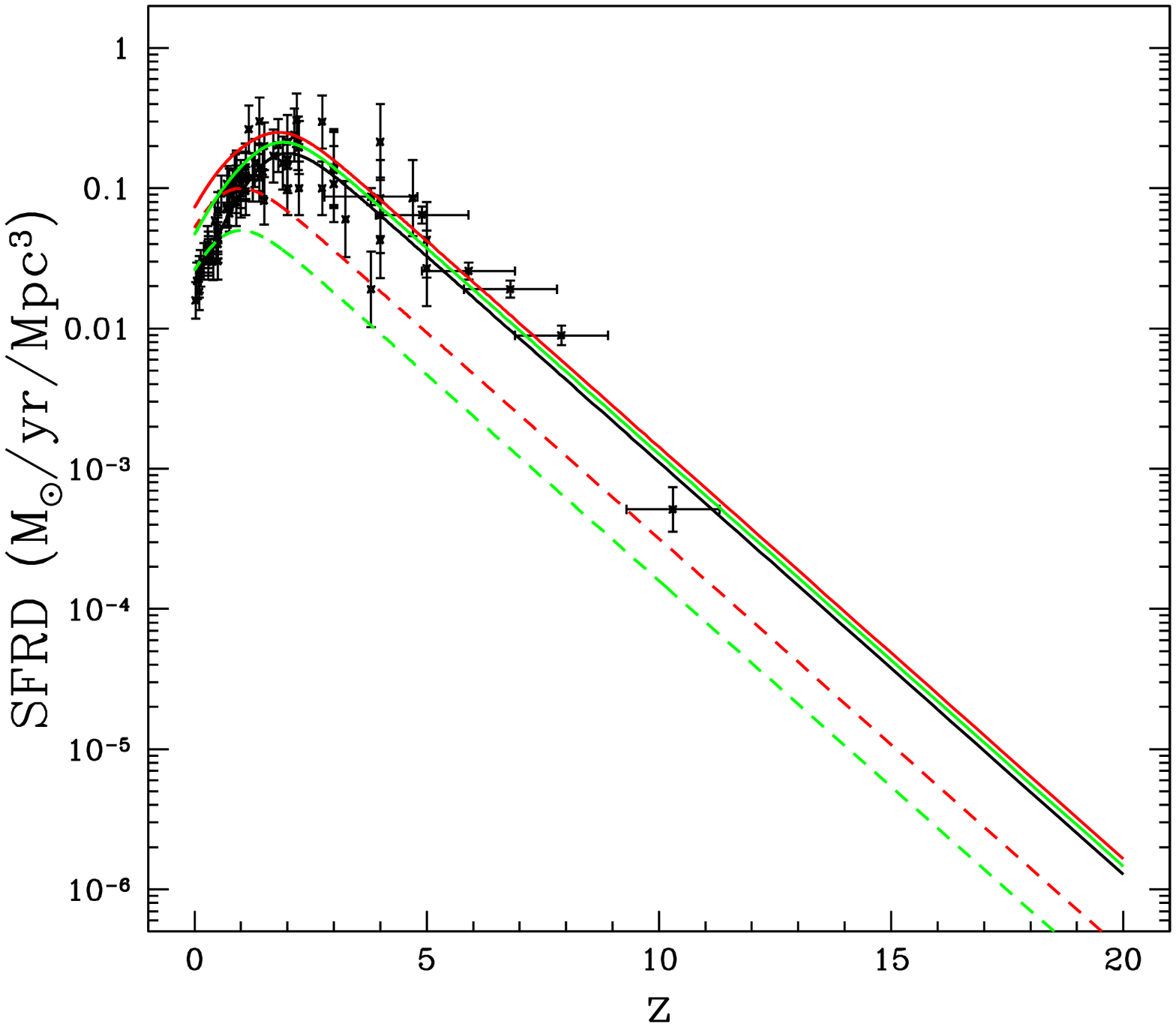, height=3in}
\vskip -1 cm
\epsfig{file=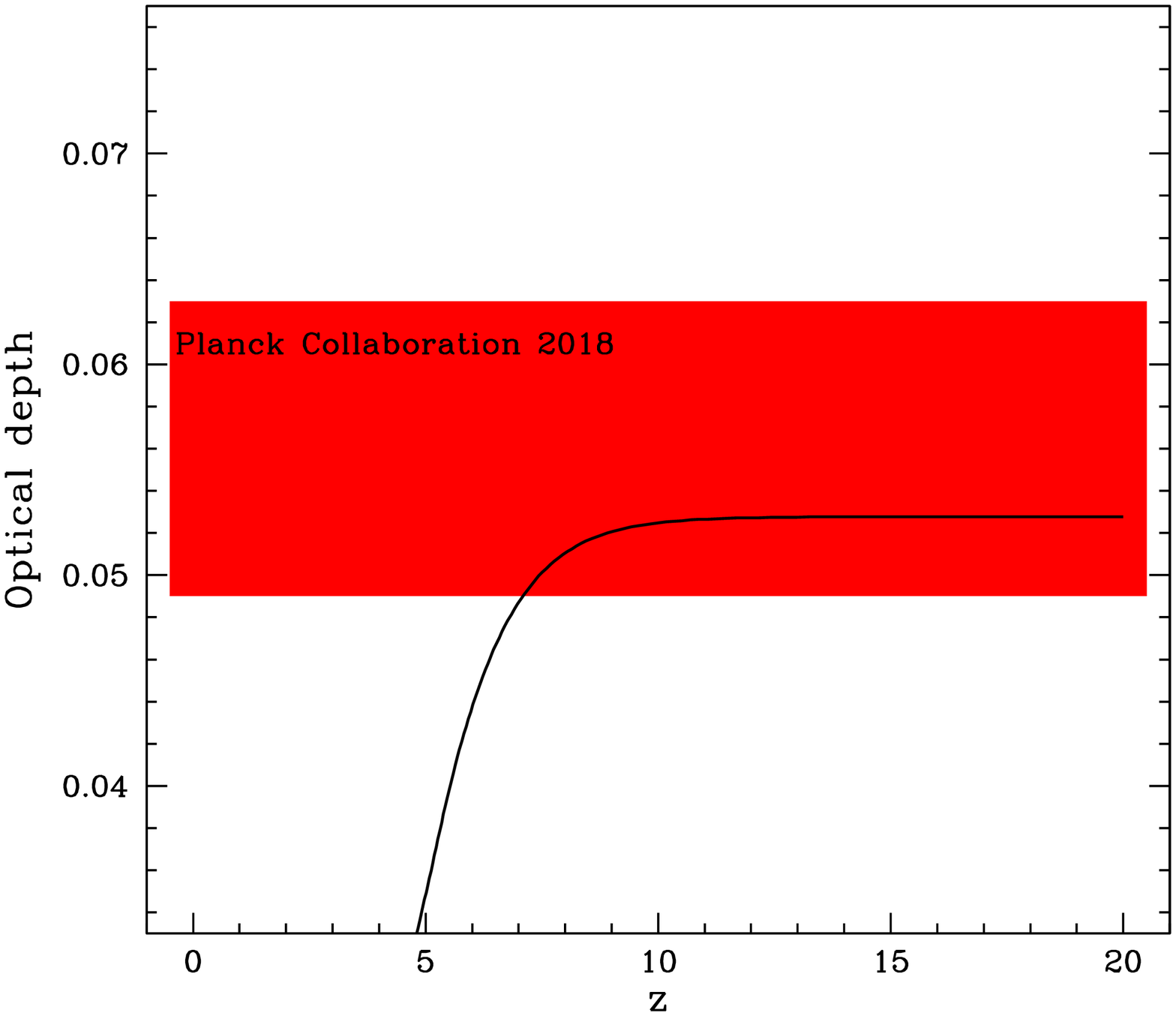, height=3in}
\end{center}
\caption{\emph{Upper panel:} 
The SFRD as a function of redshift used in our model. The black line corresponds to Model 1, whereas the solid red and green lines correspond to Model 2 with
$\nu$ = 0.1,   $\nu$ = 0.05 respectively.  The dashed red and green lines corresponds to the SFRD of the IM stellar component alone. The 
observations are taken from  \citet{Behroozi13,Bouwens14, O14,  O18}. 
\emph{Lower panel:} Evolution of the optical depth to reionization as a function of redshift. The observational constraint is indicated by a red horizontal strip \citep{Planck18}.}
\label{fig:sfr1}
\end{figure}

Fig.~\ref{fig:sfr1} (upper panel) shows the evolution of the SFRD as a function of redshift in our model, compared to observations compiled by \citet{Behroozi13} and by \citet{Bouwens14, O14, O18}. 
The black line corresponds to Model 1, i.e.,  the normal mode with no enhancement of IM stars.
The red and green dashed lines display the assumed 
SFRD of the IM mode with  $\nu$ = 0.1 (red) and  $\nu$ = 0.05 (green). 
The red and green solid lines correspond to the total SFRD (Model 2), i.e., the addition of the IM mode to Model 1.

The SFRD can also be constrained by using the optical depth to reionization, which depends on the rate of production of ionizing photons by massive stars. 
We calculate the optical depth to reionization $\tau$ as described in \citet{vangioni15}, and in particular we use the tables in \citet{schaerer02} for the number of photons produced by massive stars and assume an escape fraction of $f_{esc}=0.2$. The resulting optical depth is shown in Fig.~\ref{fig:sfr1} (bottom panel) and is  compared to the new constraints obtained from measurements of the cosmic microwave background \citep{Planck18} shown as the red band.

\section{Cosmic evolution of elements}
\label{sec:CNO}

\subsection{Evolution of Z, iron and CNO, Mg elements}

\begin{figure}
\centering
\epsfig{file=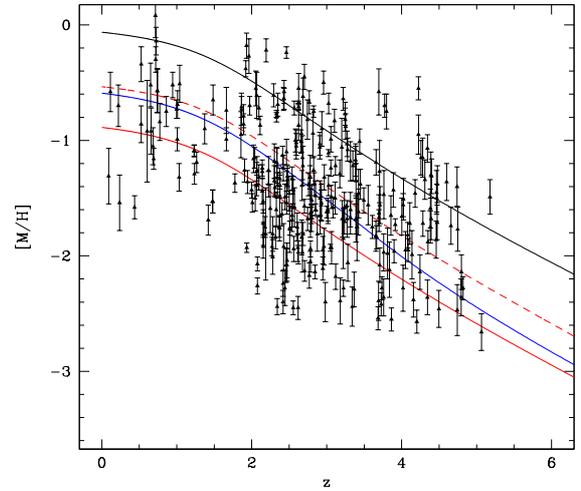, height=3in}
\caption{The evolution of [M/H] in the ISM as a function of redshift calculated using the semi-analytical model with a single sloped IMF (Model 1) for three sets of yields for massive stars. 
The black line is derived from the metallicity-dependent yields of \citet{Nomoto06}. The red lines correspond to the metallicity-dependent yields from \citet{Limongi18}. The
solid line ignores rotation and the dashed line assumes a rotational velocity of
$v_{r} = 150 $ km/s. The blue line is based on the yields from \citet{ww95}.  The yields for IM stars are taken from \citet{hg97}.  Data are taken from \citet{Raf12b}.}
\label{fig:Z}
\end{figure}

Before considering magnesium isotope evolution, we display the behavior of global metallicity and iron abundance as a function of redshift for Model 1. 
In Fig.~\ref{fig:Z},
we show the evolution of the metallicity, [M/H] $= \log($Z$/$Z$_\odot)$,  as a function of redshift for three sets of yields for massive stars with Model 1. 
The black line is derived from the metallicity-dependent yields of \citet{Nomoto06}, and the red lines (solid and dashed)  correspond to the metallicity-dependent yields from \citet{Limongi18}, without rotation and with a rotational velocity  $v_{r} = 150$ km/s, respectively. The blue line is based on the yields of \citet{ww95}. 
In all cases shown, the yields from stars with $M < 8$M$_\odot$ are taken from \citet{hg97}. As expected, the rotational effect increases the metal content in the ISM. 
As we noted above,  there is an inherent uncertainty in the resulting abundance evolution arising from the uncertainty in the 
chemical yields from massive stars. This uncertainty is clearly seen in Fig.~\ref{fig:Z} where the difference in total metallicity spans nearly an order
of magnitude. These discrepancies are the result of differing models for the microphysics of the pre-supernova evolution (such as the treatment of the convective layers) and different reaction rates.  The data, taken from \citet{Raf12b} also shows considerable dispersion. The results shown here correspond to a universal average and cannot 
account for the dispersion. The more sophisticated treatment of cosmic chemical evolution in \citet{dvorkin15} making use of chemical evolution in an inhomogeneous
background showed that indeed a dispersion in metallicity of nearly two orders of magnitude about the mean (shown here) is in fact expected.

\begin{figure}
\centering
\epsfig{file=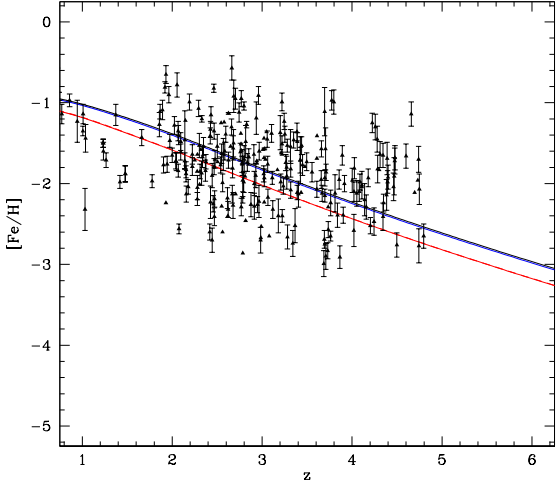, height=3in}
\caption{As in Fig.~\protect\ref{fig:Z}, the evolution of the iron abundance (relative to solar), [Fe/H], in the ISM as a function of redshift.}
\label{fig:Iron}
\end{figure}

As iron is often used as a tracer for chemical evolution, we show in 
Fig.~\ref{fig:Iron} the evolution of [Fe/H] as a function of redshift  for the same three sets of yields. 
The effect of the rotation velocity is here negligible as it affects very little the mass of the exploding iron core.
The data are again taken from \citet{Raf12b} and show significant dispersion as well which can be accounted for in an inhomogeneous model.
In what follows, we will no longer use the \citet{ww95} yields as they are found to be within the range of the other two sets of yields. 

\begin{figure}
\begin{center}
\epsfig{file=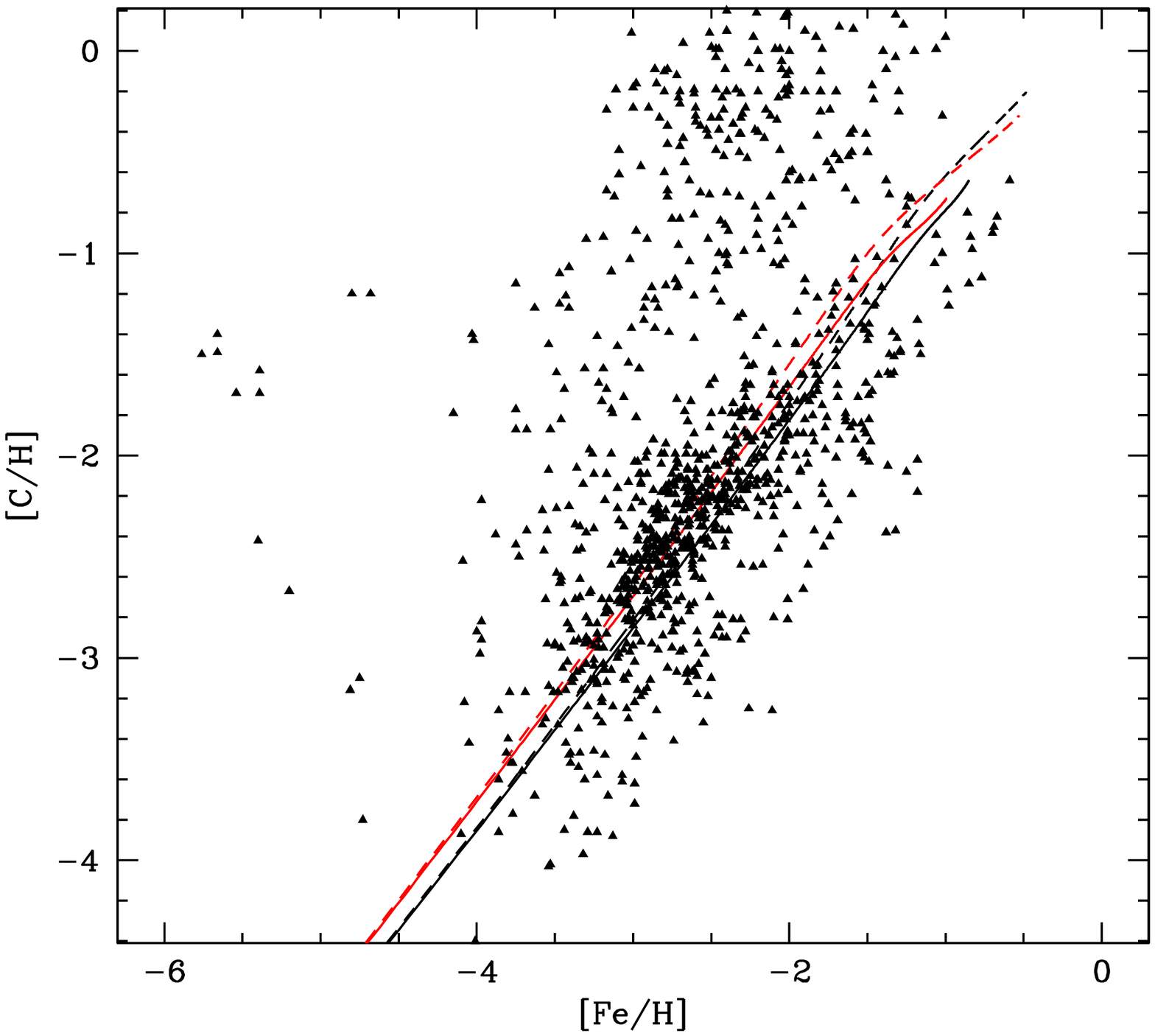, height=3in}
\vskip -1 cm
\epsfig{file=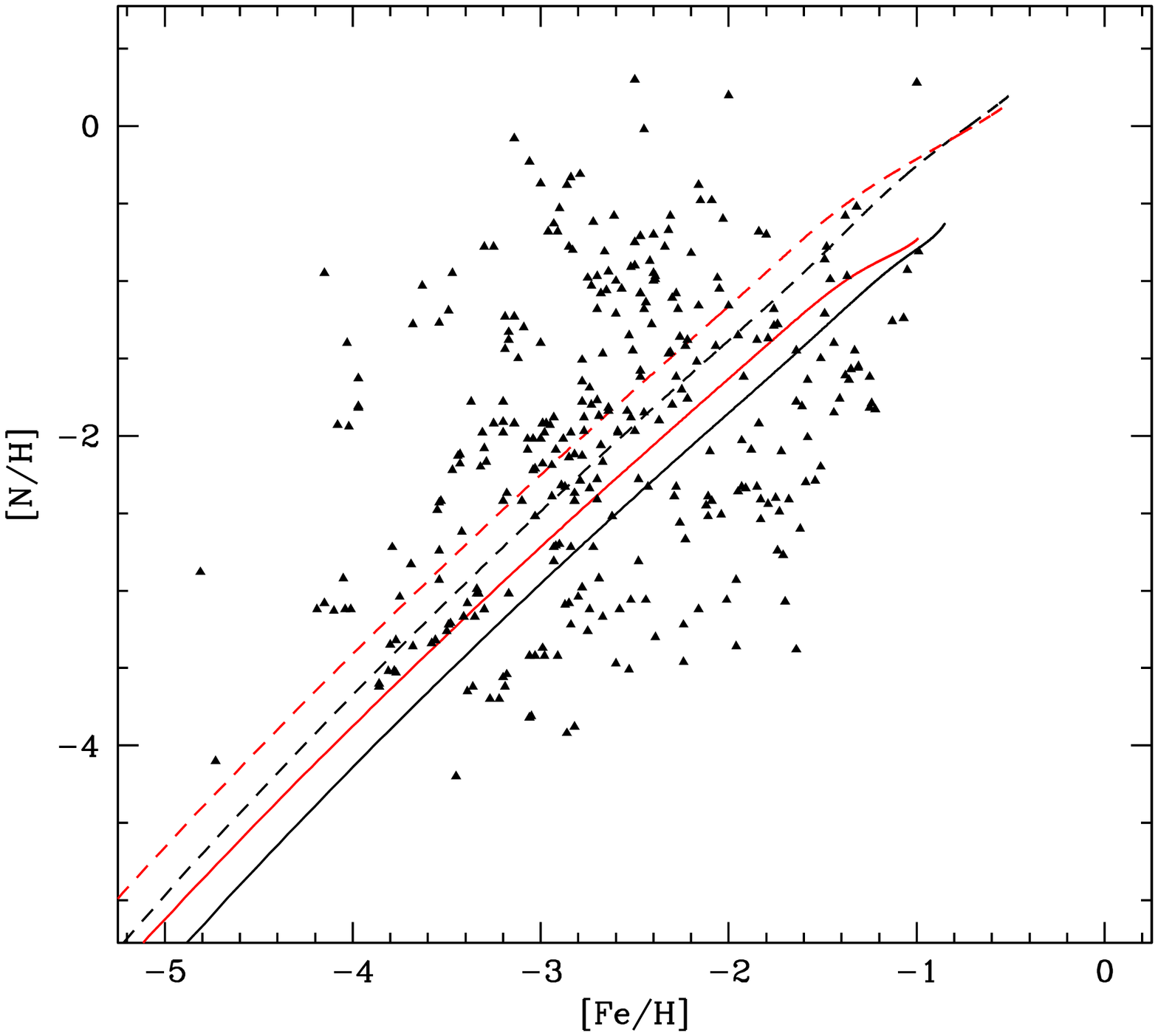, height=3in}
\vskip -1 cm
\epsfig{file=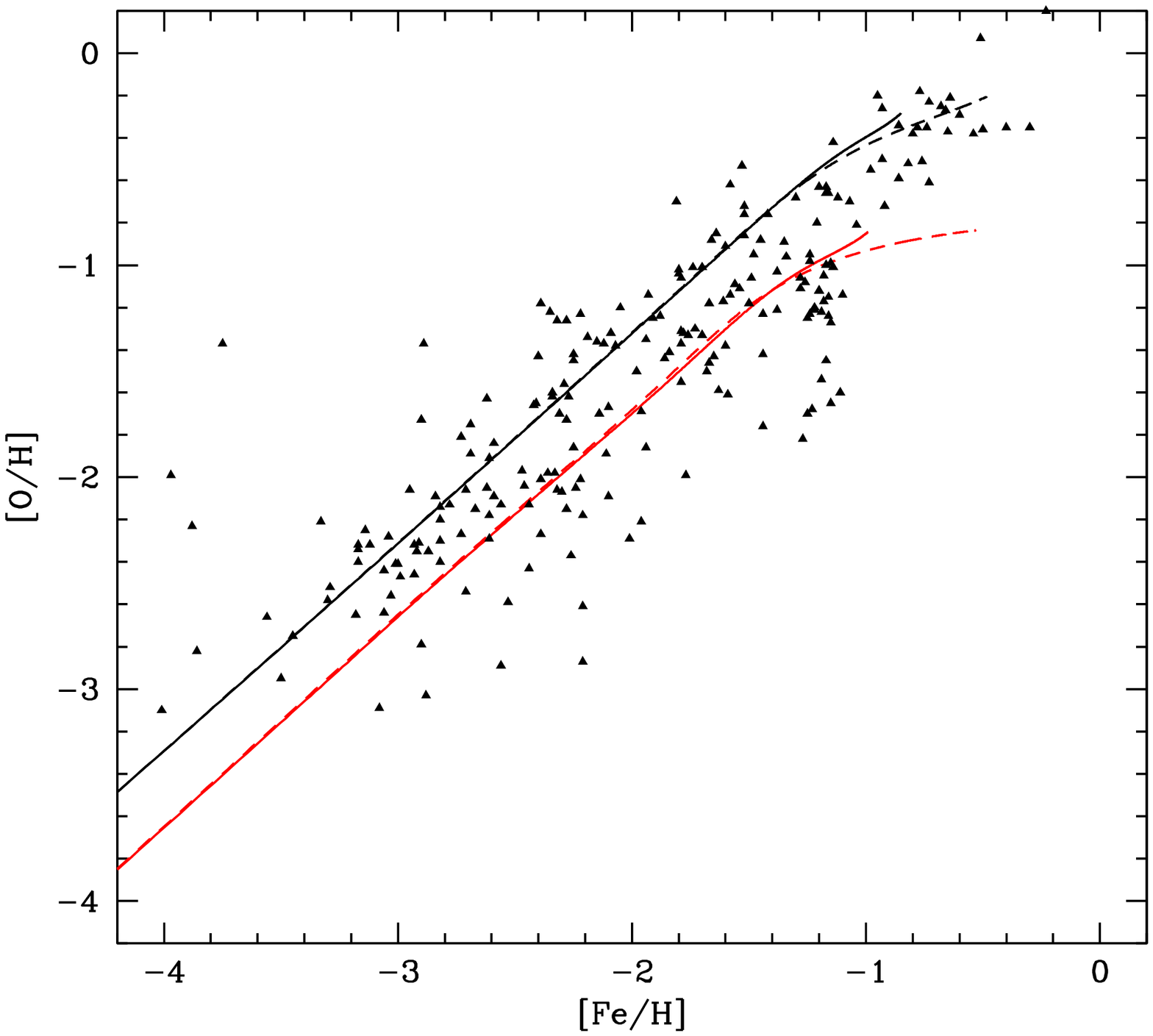, height=3in}
\end{center}
\vskip -.7 cm
\caption{The evolution of nitrogen, carbon and oxygen in the ISM as a function of [Fe/H]. The black lines are derived from the metallicity-dependent yields of \citet{Nomoto06}, and the red lines correspond to the metallicity-dependent yields from \citet{Limongi18} without rotation.
 Solid lines correspond to the single-sloped IMF (Model 1) and the dashed lines to the bimodal  IMF, with a mass range of $2 - 8$ M$_\odot$ (Model 2a).
The yields for IM stars are taken from \citet{hg97}. The observational constraints come from \citet{Cayrel04, Suda08, Suda11, Yamada13, Frebel07, Abohalima17}.}
\label{fig:CNO}
\end{figure}

\begin{figure}
\centering
\epsfig{file=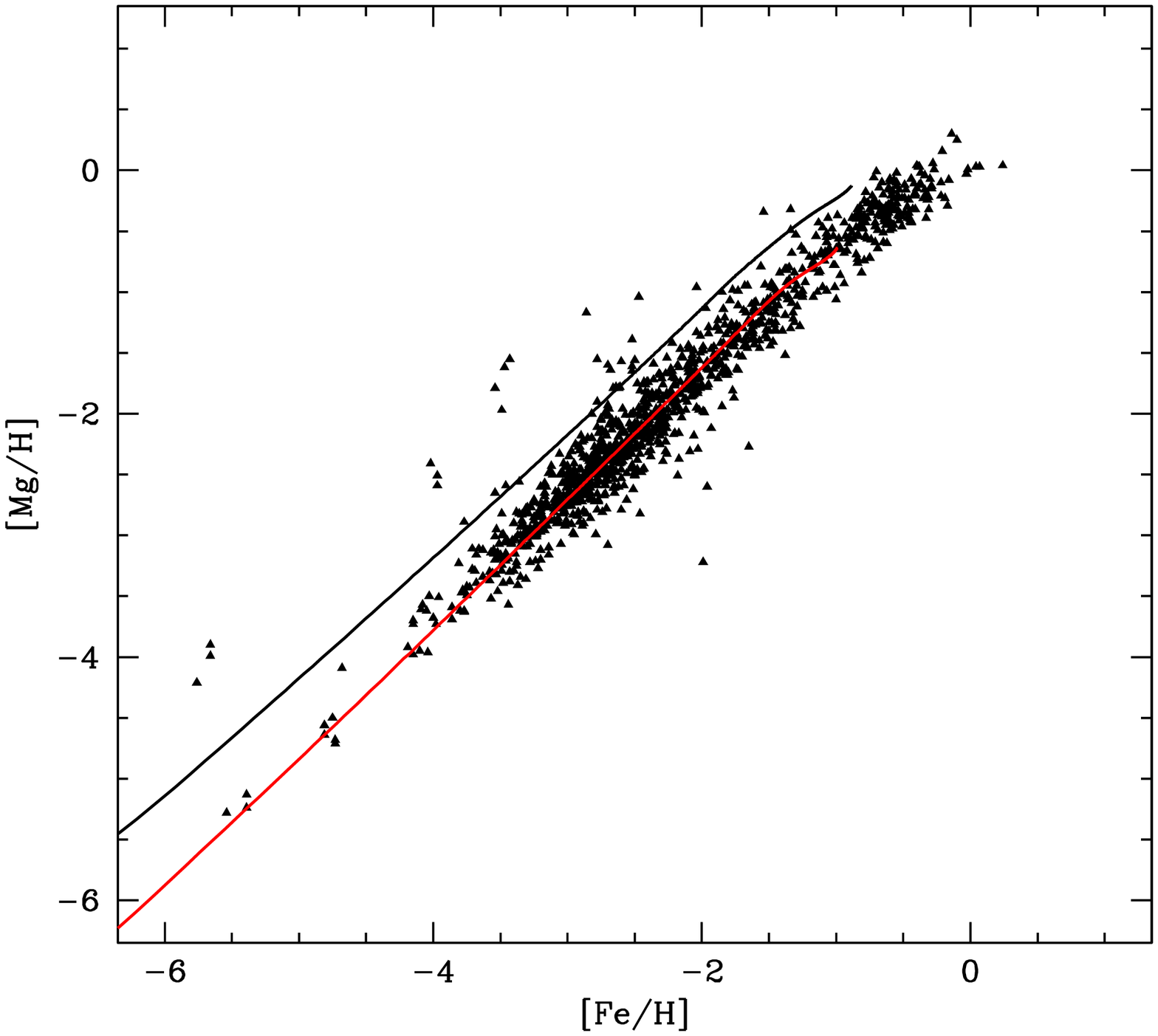, height=3in}
\caption{As in Fig. \protect\ref{fig:CNO} showing the evolution of the magnesium abundance in the ISM as a function of [Fe/H].  Observations come from \citet{Yong03a, Yong03b, Yong06, Cayrel04, Suda08, Suda11, Yamada13} \citet{Yong13}, \citet{Shetrone96}, \citet{Arnone05}, \citet{Gay00}, \citet{Fenner03}. }
\label{fig:Mg}
\end{figure}

In Fig.~\ref{fig:CNO} and Fig.~\ref{fig:Mg} we show the evolution of CNO and Mg as a function of the iron abundance. 
The solid black and red lines correspond to the two sets of yields from \citet{Nomoto06} and \citet{Limongi18} without rotation, respectively. The 
solid lines are derived using the single sloped IMF (Model 1). 
As one can see, the evolution of carbon is hardly affected by the choice of different yields.  While there is some dependence on the 
choice of yields in the nitrogen and oxygen abundances, the difference in the calculated abundance is small compared to the dispersion in the data
which are taken from multiple sources listed in the caption of each figure. The total magnesium abundance displayed in Fig.~\ref{fig:Mg} also shows some dependence
on the choice of yields which is comparable to the dispersion in the data\footnote{Throughout the paper, Mg will refer to the sum of the three Mg isotopes.}. The calculated abundances of all four elements are in good agreement with the 
bulk of the data, and due to the dispersion, we cannot use the data for these elements to discriminate between the adopted yields. 

We also show in  Figs.~\ref{fig:CNO} and~\ref{fig:Mg}, the effect of adding the intermediate mass mode with a 2 M$_\odot$ minimum mass (Model 2a). 
These results are shown by the corresponding dashed curves. As one can see, the abundances of carbon, oxygen and magnesium are virtually unaffected
by the intermediate mass mode. Nitrogen, on the other hand, does show some sensitivity and the intermediate mass mode leads to a factor of about 3
increase in N/H. This is well within the observational dispersion. Note that the magnesium abundance shown in Fig.~\ref{fig:Mg} corresponds to the
total magnesium abundance summed over all three isotopes.  As we will see in the next subsection, the abundances of the individual 
magnesium isotopes are sensitive to the inclusion of the intermediate mass mode.

\subsection{Evolution of the Mg isotopes }

In order to study the production of the magnesium isotopes in the ISM we use the yields discussed in Sections \ref{sec:massivestars} and \ref{sec:imstars} for massive stars which
undergo core collapse supernovae and intermediate mass stars with masses in the range $0.9 < M/M_\odot < 8.0$, respectively. An interpolation is made between different metallicities and masses, and tabulated values are extrapolated for masses above $40\,M_{\odot}$ when the \citet{Nomoto06} yields are used. 
Solar abundances are taken from \citet{asplund09}.

\begin{figure}
\centering
\epsfig{file=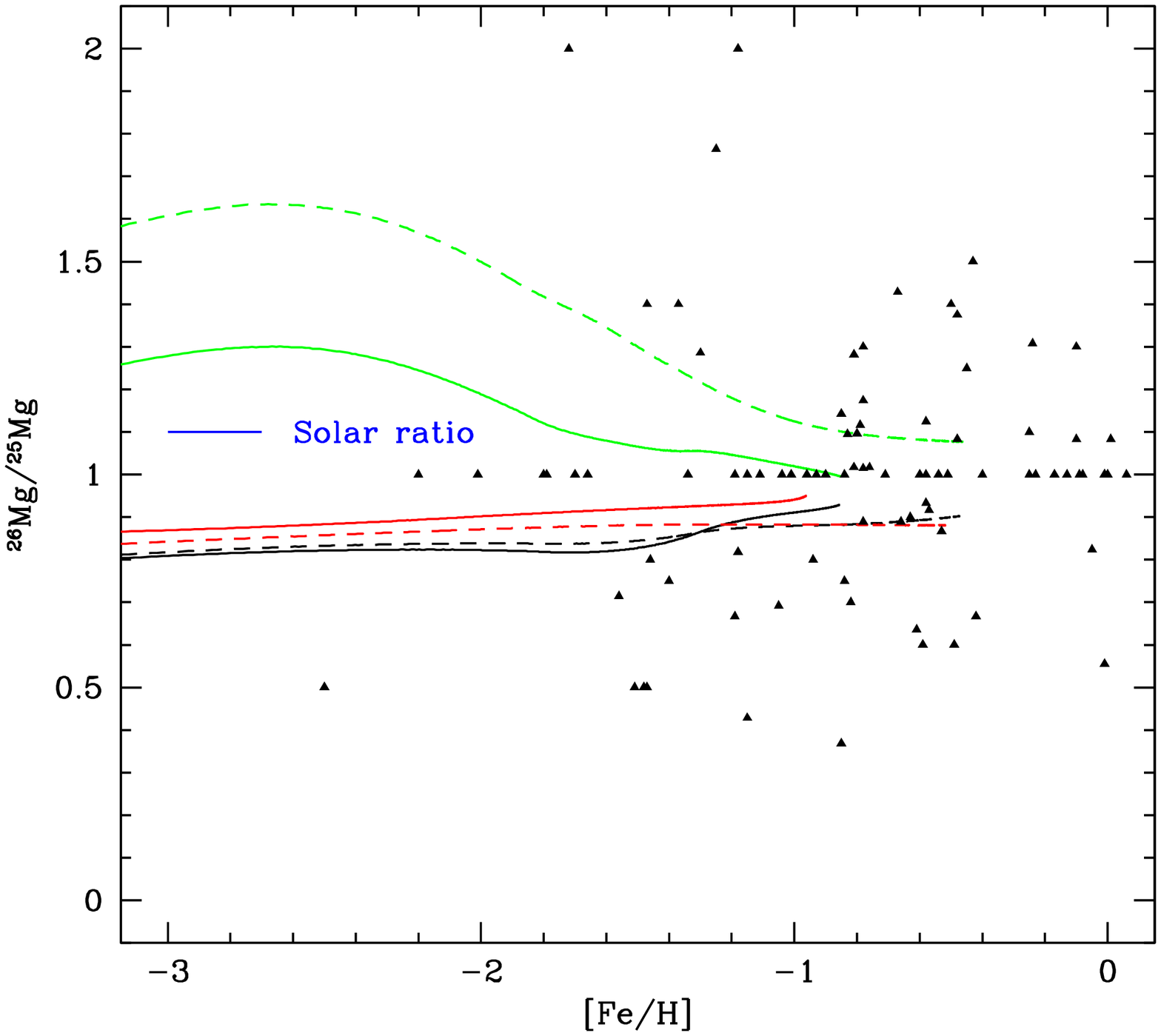, height=3in}
\caption{As in Fig.~\ref{fig:CNO} showing the evolution of the $^{26}$Mg$/^{25}$Mg ratio in the ISM  as a function of the iron
abundance. The green curves show the ratio using the yields from \citet{Karakas10} for IM stars.  
Data  are taken from \citep{Yong03a, Yong03b, Yong06, Yong13, Melendez07, Melendez09, Dacosta13,Shetrone96, Arnone05, Gay00}. }
\label{fig:Mg2625}
\end{figure}

We now compare the results of our model to the observed Mg isotope abundance measurements, focusing on the comparison between different stellar evolution models and different  sets of yields.
As in  Fig.~\ref{fig:CNO}, Fig.~\ref{fig:Mg2625} shows the evolution of the  $^{26}$Mg$/^{25}$Mg ratio in the ISM  as a function of the iron 
abundance for two sets of yields for massive stars as well as two sets of yields for IM stars all applied to Models 1 and 2a. 
Both the red and black curves use the \citet{hg97} yields for IM stars, and we find no significant difference in the ratio between the two sets of massive star models.
In contrast, the green curves employ the \citet{Karakas10} yields for IM stars (with \citet{Nomoto06} for massive stars). Here we see an enhancement in $^{26}$Mg. 
Not surprisingly, there is further enhancement of $^{26}$Mg in this case when the additional IM mode (Model 2a) is included. 
Overall, the ratio of the two heavier isotopes is relatively fixed and does not depend sensitively on the addition of an intermediate mass mode.

\begin{figure}
\centering
\epsfig{file=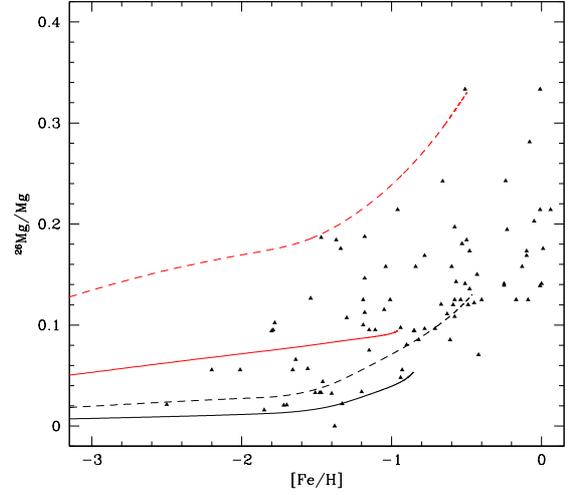, height=3in}
\caption{As in Fig.~\ref{fig:CNO} showing the evolution of the $^{26}$Mg/Mg ratio  in the ISM as a function of the iron abundance.}
\label{fig:Mg26Mg1}
\end{figure}

\begin{figure}
\centering
\epsfig{file=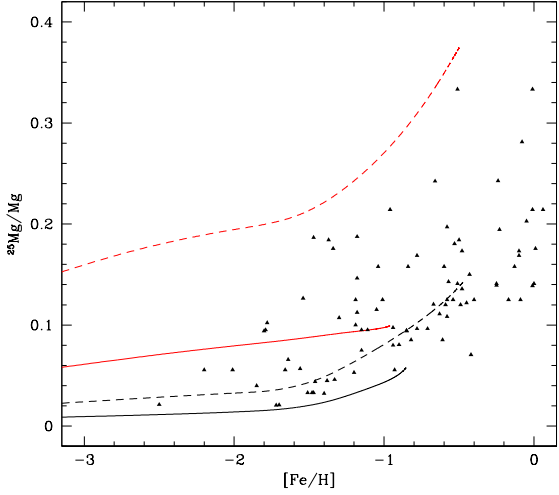, height=3in}
\caption{As in Fig.~\ref{fig:CNO} showing the evolution of the $^{25}$Mg/Mg ratio.}
\label{fig:Mg25Mg1}
\end{figure}

\begin{figure}
\centering
\epsfig{file=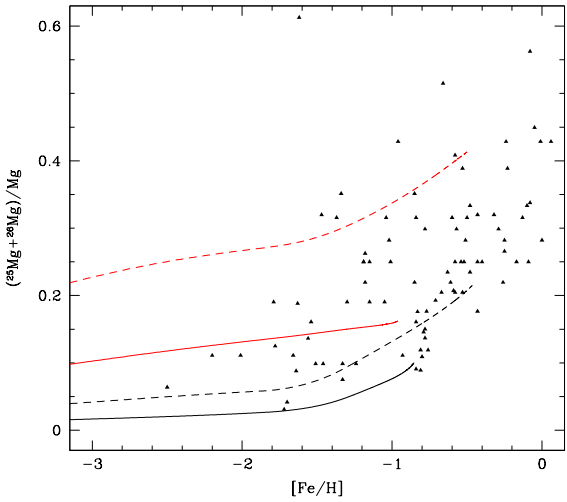, height=3in}
\caption{As in Fig.~\ref{fig:CNO} showing the evolution of the ($^{25}$Mg + $^{26}$Mg)/Mg ratio.}
\label{fig:Mg2625Mg241}
\end{figure}

In contrast, we see in  Figs.~\ref{fig:Mg26Mg1} -- \ref{fig:Mg2625Mg241} very significant differences between the models for the ratios
of $^{26}$Mg to the total Mg abundance (Fig.~\ref{fig:Mg26Mg1}), $^{25}$Mg to the total Mg abundance (Fig.~\ref{fig:Mg25Mg1}), and 
the sum of  $^{25}$Mg and  $^{26}$Mg relative to the total abundance of  Mg (Fig.~ \ref{fig:Mg2625Mg241}). Once again,  solid lines (black lines corresponding to the
yields of \citet{Nomoto06} and  red lines  from \citet{Limongi18}) show a difference by a factor of a few. 
In both cases, the yields for IM stars are taken from \citet{hg97}.  Had we used the \citet{Karakas10} yields (with \citet{Nomoto06} for massive stars),
results would appear very similar to the black curves shown.  That is, the ratios to total Mg are not particularly sensitive to the choice of IM yields. 
The dashed lines show results for the isotope ratios when the intermediate
mass mode with masses between 2 and 8 M$_\odot$ (Model 2a) is included.
Using the \citet{Nomoto06} yields, it is clear that adding the IM mode allows for a better fit to the observations. 
However, the  \citet{Limongi18}) yields synthesize significantly less $^{24}$Mg (see Table \ref{table1} and Fig.~\ref{fig:Mg}) 
and adding the IM mode would lead to an excess in these
isotopes compared to the observations.
Note that the evolutionary tracks for each model reach a different value of [Fe/H],  and in no model is solar iron achieved.
This is because, as noted earlier, we are showing cosmic averages of the element abundances in the ISM. In an inhomogeneous model such as that described in
\citet{dvorkin15}, some regions (in the universe rather than in the Galaxy) achieve solar (and supersolar) metallicities, but on average, they do not.
Data come from different sources \citep{Yong03a, Yong03b, Yong06, Yong13, Melendez07, Melendez09, Dacosta13,Shetrone96, Arnone05, Gay00}.

\begin{figure}
\centering
\epsfig{file=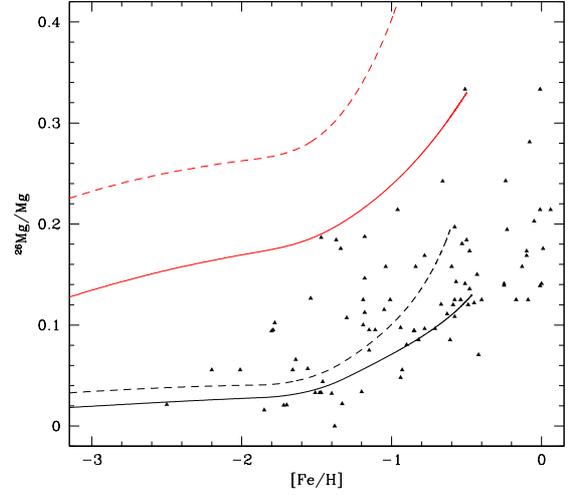, height=3in}
\caption{ As in Fig.~\ref{fig:Mg26Mg1} showing the evolution of  $^{26}$Mg/Mg in the ISM as a function of the iron abundance
comparing Models 2a and 2b. The black lines are derived from the metallicity-dependent yields of \citet{Nomoto06}, and the red lines correspond to the metallicity-dependent yields from \citet{Limongi18} without rotation. Solid lines correspond to the  
bimodal model with the mass range: $2 - 8$ M$_\odot$  (Model 2a) and the dashed lines correspond to the $5 - 8$ M$_{\odot}$ IM mass range (Model 2b).}
\label{fig:Mg26Mg2}
\end{figure}

\begin{figure}
\centering
\epsfig{file=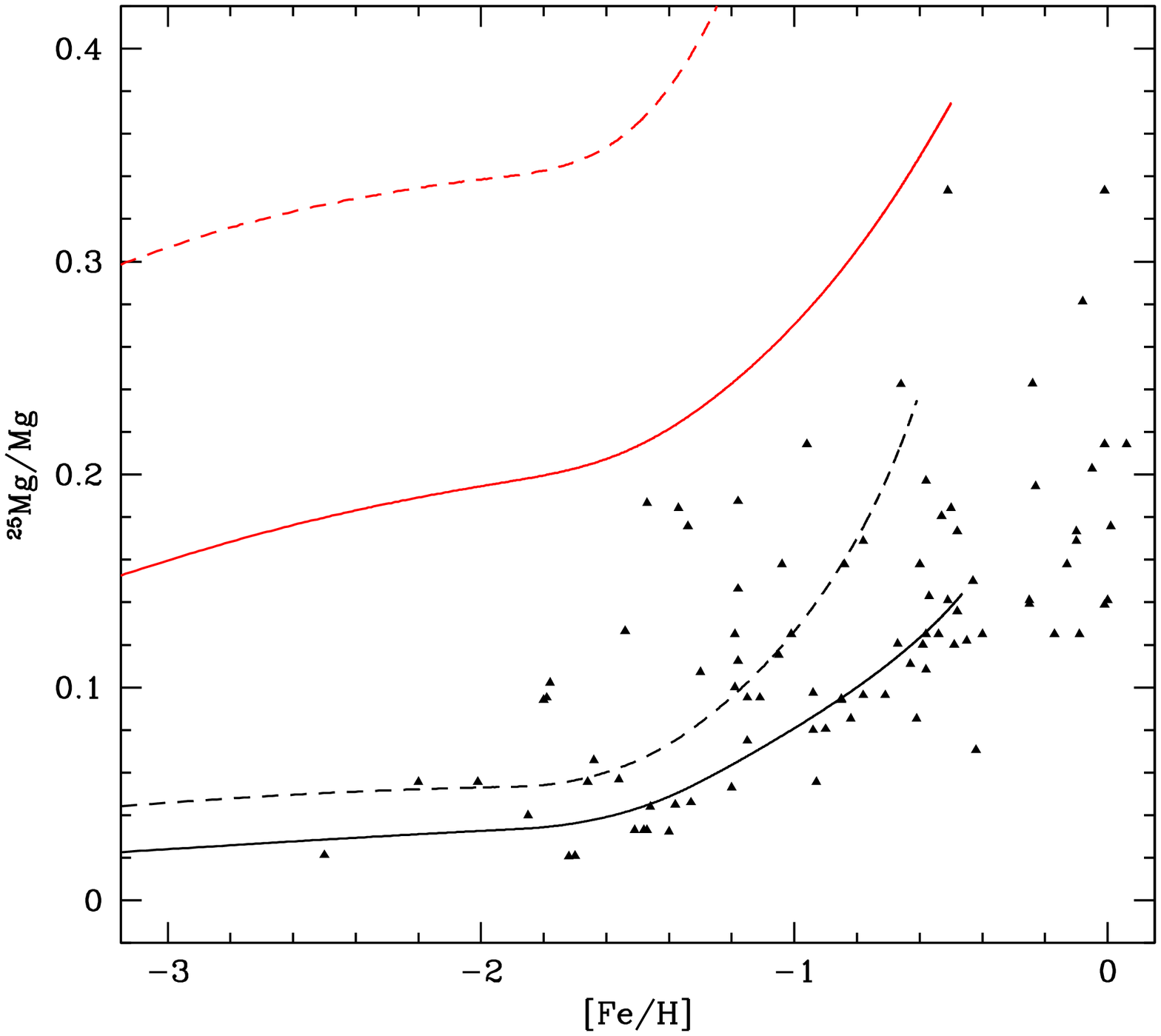, height=3in}
\caption{As in Fig.~\ref{fig:Mg26Mg2} showing the evolution of $^{25}$Mg/Mg  in the ISM comparing Models 2a and 2b.}
\label{fig:Mg25Mg2}
\end{figure}

\begin{figure}
\centering
\epsfig{file=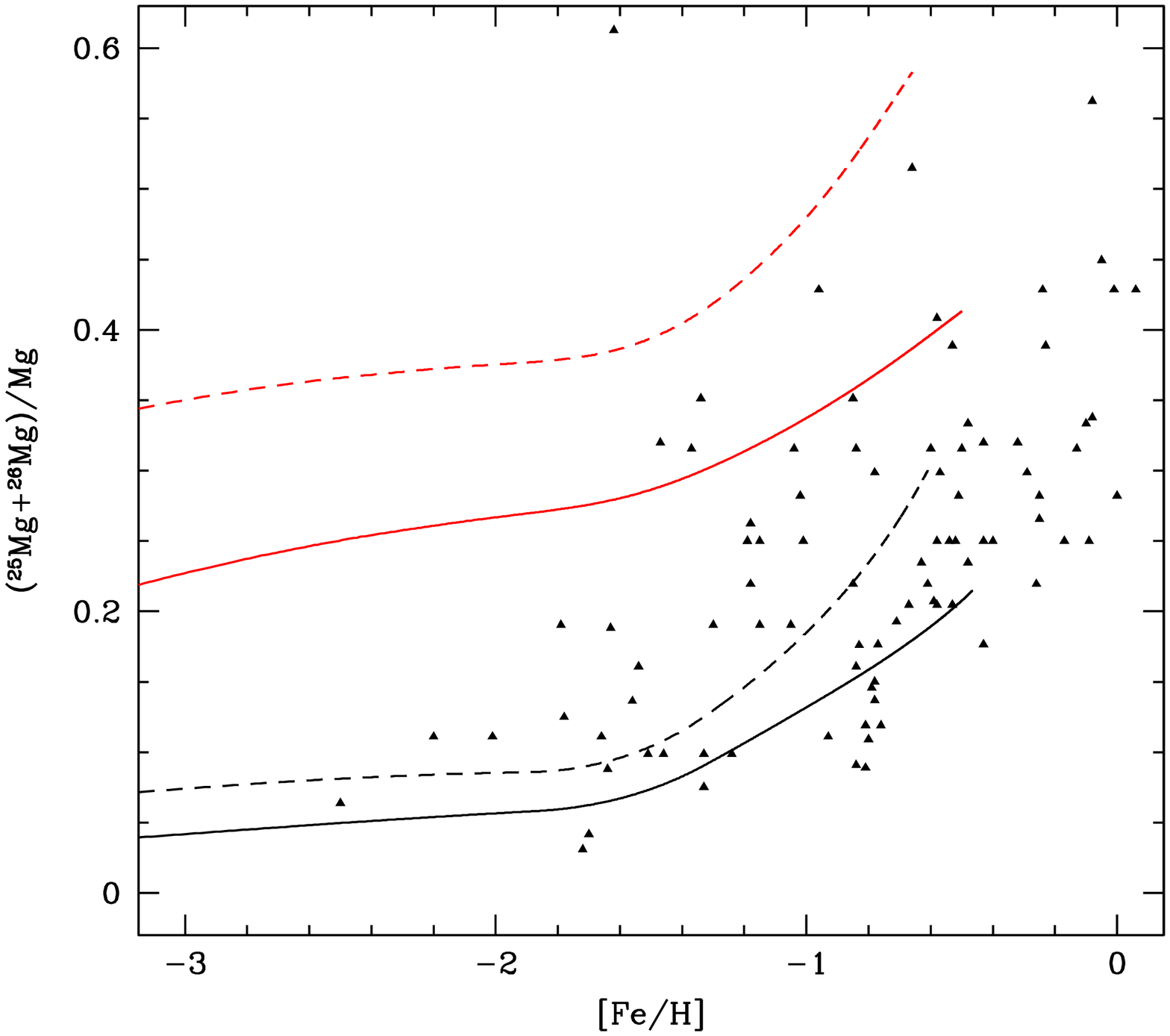, height=3in}
\caption{As in Fig.~\ref{fig:Mg26Mg2} showing the evolution of ($^{25}$Mg + $^{26}$Mg)/Mg  in the ISM comparing Models 2a and 2b.}
\label{fig:Mg2625Mg242}
\end{figure}

Similarly, Figs.~\ref{fig:Mg26Mg2} -- \ref{fig:Mg2625Mg242}  show the impact of the lower mass limit, $m_{inf}$, of the IM mode. 
In these figures we have increased $m_{inf}$ from 2 to 5 $M_{\odot}$ (Model 2b). In these figures, we are comparing
Model 2a and 2b, so all results include the addition of the IM mode albeit with different lower mass limits. Solid curves correspond
to Model 2a and are identical the dashed curves in Figs.~\ref{fig:Mg26Mg1} -- \ref{fig:Mg2625Mg241}. Dashed curves here correspond to 
Model 2b which increases the heavy isotope ratios. This leads to further improvement of the model predictions 
using the \citet{Nomoto06} yields with respect to the observations. The IM mode with the yields of \citet{Limongi18} do not fit the observations,
but this may be a result of the decreased $^{24}$Mg abundance produced in these models, rather than an excess in the heavy isotopes.

\begin{figure}
\begin{center}
\epsfig{file=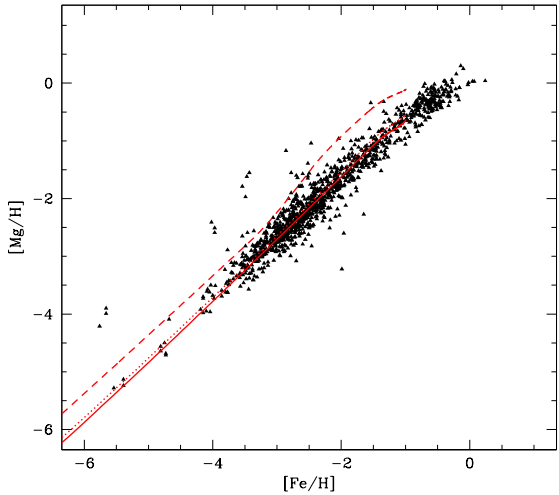, height=3in}
\end{center}
\caption{The evolution of  the total magnesium abundance in the ISM as a function of the iron abundance. Red lines corresponding to metallicity-dependent yields from \citet{Limongi18} with three different choices of rotation velocity, $v_r$ for Model 1. Solid, dotted and dashed lines correspond to $v_r = 0$ km/s,  $v_r = 150$ km/s, $v_r = 300$ km/s, respectively.}
\label{fig:Mgventa}
\end{figure}

Finally, it is interesting to analyze the impact of rotation in massive stars on the evolution of the Mg isotope ratios. In \citet{Limongi18} three sets of yields are given for three different rotation velocities:  $v_r = 0$ km/s (solid line),  $v_r = 150$ km/s (dotted line), $v_r = 300$ km/s (dashed line). 
In Fig.~\ref{fig:Mgventa}, we show the evolution of the total Mg abundance as a function of [Fe/H] for Model 1. 
The red solid curve here, is the same as that given in Fig.~\ref{fig:Mg}.
As one can see, while rotation lowers the Mg abundance, the effect on the total Mg abundance is rather small.
As we already saw in Fig.~\ref{fig:Mg}, adding the IM component makes an imperceptible difference to the total Mg abundance,
and the effect of rotation on Model 2 is the same. That is, the evolution of Mg vs. [Fe/H] using the \citet{Limongi18} yields looks identical for both Models 1 and 2. 

\begin{figure}
\begin{center}
\epsfig{file=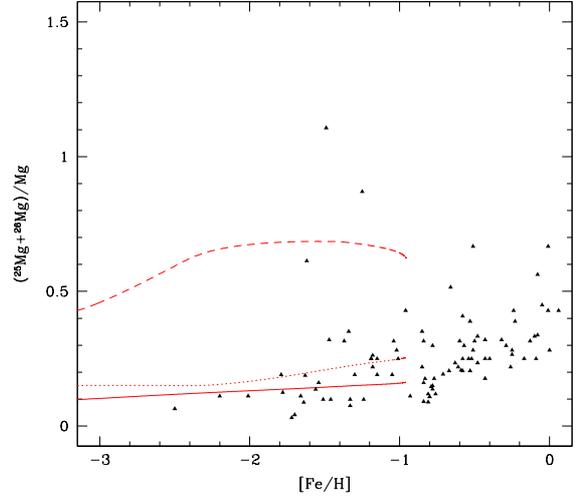, height=3in}
\end{center}
\caption{As in  Fig.~\ref{fig:Mgventa}, the evolution of ($^{25}$Mg + $^{26}$Mg)/Mg ratio as a function of the iron abundance. }
\label{fig:Mg2625Mg24venta}
\end{figure}

In contrast, rotation has a large effect on the isotopic ratios of Mg. In 
Fig.~\ref{fig:Mg2625Mg24venta}, we show the isotopic evolution for Model 1. 
When the rotational velocity is relatively low, the predicted ratios are in 
relatively good agreement with observations. At higher rotational velocities,
the yields of the heavy isotopes are strongly dependent on the stellar mass.
At the relatively low masses of 13 and 15 M$_\odot$ and metallicities [Fe/H] = -2 and -1,  the  yield of ${}^{26}$Mg 
shows a dramatic rise at $v_r = 300$ km/s with ${}^{26}$Mg/${}^{24}$Mg $>$ 1 (M. Limongi, private communication).   
Because of the power-law dependence of the IMF, the lower mass end of the IMF for massive stars 
dominates the chemical evolution and when integrated over the IMF leads to a large enhancement of ${}^{26}$Mg.
This is clearly seen in  Fig.~\ref{fig:Mg2625Mg24venta} where there is a substantial increase in the heavy isotope ratio when 
the highest rotation velocity is considered, and $v_r = 300$ km/s  shows a behavior  well outside of the bulk of data
and should be excluded if all stars are assumed to have the highest rotation velocities considered.

\section{Conclusion}
\label{sec:conclusion}

Our knowledge of the detailed history of star formation both in the Galaxy and on a cosmic scale relies on 
stellar nucleosynthesis and our ability to trace the evolution of the chemical abundances from the 
formation of the first stars at zero metallicity to the present day at solar metallicity.  
Different elements and isotopes reveal different aspects of the stellar history. For example,
the production of the $\alpha$ elements is strongly tied to the their production in 
massive stars ($M > 8$ M$_\odot$) ending as core-collapse supernovae.
Other elements such as iron, span a combination of origins including both type I and II supernovae.
The relative importance of intermediate mass stars can be gleaned from the abundances of elements such as
nitrogen and as we have argued here, the heavy isotopes of Mg.

Here, we have worked in the context of a model of cosmic chemical evolution.
The model is constrained by the observed star formation rate density obtained from the luminosity function
measured at high redshift. The optical depth to reionization derived from CMB observations also constrains 
the rate of star formation at high redshift. While such models necessarily carry large uncertainties 
(due to the choice of the IMF, as well as uncertainties in the calculated stellar yields),
general evolutionary tendencies can be extracted from them. As we have seen, the overall
evolution of iron and the $\alpha$ elements such as C and O is well reproduced in these models.
Indeed the dispersion in the data at low metallicity and/or high redshift is far greater than the 
spread in model predictions. Recall that the type of evolutionary model we have employed 
calculates only average abundances, and a more detailed model such as that based on 
merger trees is needed to understand the degree of dispersion observed in the data \citep{dvorkin15}.

In this paper we have explored the evolution of the magnesium isotopic 
abundances in the ISM using our model of cosmic chemical evolution based on hierarchical structure formation.
The abundances of the heavier Mg isotopes are primarily produced in intermediate mass stars and 
therefore the abundance ratios of these isotopes provide insight into the relative importance of
intermediate mass stars in chemical evolution. 
It is interesting to note that the apparent conflict between the EBL density and IR measurements may also
imply a need for an additional component of IM stars \citep{fardal,2018arXiv180805208C}.

Taking into account the inherent uncertainty introduced by the choice
of stellar yields, we have explored several sets of nucleosynthetic yields. As expected, we confirm that the 
bulk of $^{24}$Mg in the ISM is produced by massive stars.  However, 
a single sloped Salpeter-like IMF does not reproduce the observed evolutionary behavior of  $^{25}$Mg and $^{26}$Mg 
which show enhancements at later metallicities.  
Instead, an additional component (making the IMF bimodal) of intermediate mass stars seems to be required to fit the observational constraints on the isotopic ratios. This conclusion holds independently of the choice of yields.
However, the relative importance of the IM component is very sensitive to the choice of yields from
IM stars.  Adding to the uncertainty in the heavy isotopic yields is the degree of rotational velocity which can strongly
affect the yield of $^{26}$Mg.

It is clear that there is a strong interplay between observations of element (and isotopic) abundances, 
calculations of nucleosynthetic yields and the modeling of the chemical history of these abundances.
Progress in any one of these three areas relies on progress in the other two.
Here, we have argued that progress on evolutionary models and the relative importance of intermediate mass
stars relies on high precision data, and a better understanding of the nuclear processing in the AGB phase
of intermediate mass stars.

\section*{Acknowledgements}
We would like to thank Marco Limongi for useful correspondence concerning the effects of stellar rotation on the production of the heavy isotopes.
This work is made in the ILP LABEX (under reference ANR-10-LABX-63) supported by French state funds managed by the ANR within the Investissements d'Avenir programme under reference ANR-11-IDEX-0004-02. 
The work of  K.A.O. was supported in part by DOE grant DE--SC0011842 at the University of Minnesota. 

\bibliographystyle{mnras}
\bibliography{Magnesium}
\label{lastpage}
\end{document}